\documentclass[12pt]{article}
\usepackage{a4wide}
\usepackage{epsfig}
\usepackage{amsmath}

\newlength{\absize}
\catcode`@=11
\def\citer{\@ifnextchar [{\@tempswatrue\@citexr}{\@tempswafalse\@citexr[]}}

\def\@citexr[#1]#2{\if@filesw\immediate
  \write\@auxout{\string\citation{#2}}\fi
  \def\@citea{}\@cite{\@for\@citeb:=#2\do
    {\@citea\def\@citea{--\penalty\@m}\@ifundefined
       {b@\@citeb}{{\bf ?}\@warning
       {Citation `\@citeb' on page \thepage \space undefined}}%
\hbox{\csname b@\@citeb\endcsname}}}{#1}}
\catcode`@=12

\begin{document}
  \thispagestyle{empty}
  \pagestyle{empty}
  \renewcommand{\thefootnote}{\fnsymbol{footnote}}
\newpage\normalsize
    \pagestyle{plain}
    \setlength{\baselineskip}{4ex}\par
    \setcounter{footnote}{0}
    \renewcommand{\thefootnote}{\arabic{footnote}}
\newcommand{\preprint}[1]{%
  \begin{flushright}
    \setlength{\baselineskip}{3ex} #1
  \end{flushright}}
\renewcommand{\title}[1]{%
  \begin{center}
    \LARGE #1
  \end{center}\par}
\renewcommand{\author}[1]{%
  \vspace{2ex}
  {\Large
   \begin{center}
     \setlength{\baselineskip}{3ex} #1 \par
   \end{center}}}
\renewcommand{\thanks}[1]{\footnote{#1}}
\begin{flushright}
\end{flushright}
\begin{center}
\end{center}
\begin{flushright}
\end{flushright}
\vskip 0.5cm

\begin{center}
{\large \bf Heisenberg Quantization in Physical Space Based on
Established Experiments}
\end{center}
\vspace{1cm}
\begin{center}
Jian-Zu Zhang$\;^{\ast}$
\end{center}
\vspace{1cm}
\begin{center}
Institute for Modern Physics, Box 316, East China University of
Science and Technology, Shanghai 200237, P. R. China
\end{center}
\vspace{1cm}


\begin{abstract}
It is clarified that Heisenberg quantization was proposed in empty
space.
Based on established experiments, the generalized Heisenberg
quantization in physical space is obtained.
Physical space quantization includes important new physics:
Proving that physical space is noncommutative space;
 %
Exploring the existence of a non-zero minimal length scale, which
leads to new space structures and the existence of the space minimal
finite volume;
Finding a new correlativity of the property of space with the motion
status of the system: space non-commutativity is determined by the
momentum non-commutativity.

\end{abstract}

\begin{flushleft}
$^{\ast}$ E-mail address: jzzhang@ecust.edu.cn

\end{flushleft}

\hspace*{1.3cm} Correlativity of space property with motion status

\clearpage

Quantum mechanics based on Heisenberg quantization ({\bf HQM}) is
one of the most successful physical theories. It has obtained highly
accurate test of all experiments.
It is the theoretical basis of modern physics, chemistry,
information technology and so on.
Heisenberg quantization was established in empty space (pure
vacuum).
It reads (we consider two dimensional space)
\begin{equation}
\label{Eq:xp}
[x_{i},p_{j}]=i\hbar\delta_{ij}, [x_{i},x_{j}]=0,
[p_{i},p_{j}]=0,(i,j=1,2),
\end{equation}
where it is supposed that both position-position and
momentum-momentum are commuting.
Eq.(\ref{Eq:xp}) is the basis of {\bf HQM} (Tree equations all
together in Eq.(\ref{Eq:xp}) is also called Heisenberg-Weyl
algebra).

In the real universe, space and the cosmic magnetic field are
simultaneously exist in a very large cosmic scale since the early
Universe. In the reality, empty space does not exist; Only physical
space (space filled by the cosmic magnetic field) exists. The cosmic
magnetic field has a role of the intrinsic magnetic field of
physical space.
The present intergalactic magnetic field $\bf B^{c}$ has an
intensity of
\begin{equation}
\label{Eq:Bc}
\bf B^{c}\sim 10^{-8} Gauss=10^{-12} T
\end{equation}
and a dominant scale-length $\sim$ 10 kpc~\cite{Harr}.
In the following, we mean that the cosmic magnetic field is this
$\bf B^{c}$.
Although $\bf B^{c}$ is so weak, its role in quantum mechanics is
much more important than our previous understanding. In the
traditional way of {\bf HQM}, $\bf B^{c}$ is treated as perturbation
in the Schr\"odinger equation.
However, there is another way: this intrinsic magnetic field $\bf
B^{c}$, according to the minimal coupling of the
electromagnetic interaction, is included in the definition of
mechanical momentum, it follows that important new physics emerge,
which are not noticed before.

When the cosmic magnetic field $\bf B^{c}$ is considered, there are
two possibilities for generalizing Heisenberg quantization to
physical space:

1) Both commutators of the momentum-momentum and the
position-position are non-commuting. In this case, the generalized
Heisenberg quantization ({\bf GHQ}) reads
\begin{equation}
\label{Eq:tilde-xp}
[\widetilde {x}_i,\widetilde {p}_j]=i\hbar\delta_{ij}, [\widetilde
{p}_i,\widetilde {p}_j]=i\xi^2\eta_c\epsilon_{ij}, [\widetilde
{x}_i,\widetilde {x}_j]=i\xi^2\theta_c\epsilon_{ij},
(i,j=1,2),
 %
\end{equation}
where ($\widetilde {\bf p},\widetilde {\bf x}$) are expressed by
(${\bf x},{\bf p}$):
\begin{equation}
\label{Eq:tilde-p,x}
\widetilde {p}_i= \xi[ p_i+\eta_c\epsilon_{ij}x_j/(2\hbar)],
\widetilde {x}_i= \xi[x_i-\theta_c\epsilon_{ij}p_j/(2\hbar)].
\end{equation}
In the above, $\epsilon_{ij}$ is an antisymmetric unit tensor,
$\epsilon_{12}=-\epsilon_{21}=1,$ $\epsilon_{11}=\epsilon_{22}=0;$
$\xi=[1+\theta_c\eta_c/(4\hbar^2)]^{-1/2}$ is the scaling factor ,
which is demanded by the consistence of Eq.(\ref{Eq:tilde-xp}).
$\eta_c$ and $\theta_c$ are, respectively, the momentum-momentum
noncommutative parameter and the position-position noncommutative
parameter.

In the tilde system ($\widetilde {\bf p},\widetilde {\bf x}$), Heisenberg
quantization [the first equation of Eq.(\ref{Eq:xp})] is maintained.

2) Only the commutator of the momentum-momentum is non-commuting,
but the commutator of the position-position is commuting. In this case,
$\theta_c=0$, thus $\xi=1$, $\widetilde x_i=x_i$, then
Eq.(\ref{Eq:tilde-xp}) and~Eq.(\ref{Eq:tilde-p,x}) reduce to
\begin{equation}
\label{Eq:tilde-xp-2}
[\widetilde {x}_i,\widetilde {p}_j]=i\hbar\delta_{ij}, [\widetilde
{p}_i,\widetilde {p}_j]=i\eta_c\epsilon_{ij}, [\widetilde
{x}_i,\widetilde {x}_j]=0,\\
\nonumber
\end{equation}
\begin{equation}
\label{Eq:tilde-xp-2}
\widetilde {p}_i=  p_i+\eta_c\epsilon_{ij}x_j/(2\hbar),
\widetilde {x}_i=x_i, (i,j=1,2),
 %
 %
\end{equation}
which also maintains Heisenberg quantization.

{\bf Proof of Eq.(\ref{Eq:tilde-p,x})}. When $\bf B^{c}$ is
considered, the minimal coupling of the electromagnetic interaction
demands that the canonical momentum ${\bf p}$ should be replaced by
$\widetilde {p}_i\sim p_i-q A_i^{c}(x)$, where $\bf A^{c}$ is the
vector potential of $\bf B^{c}$. At the micro-scale, $\bf B^{c}$ can
be considered as a constant field. Taking the direction of $\bf
B^{c}$ as the $z$ direction, it follows that
$A_i^{c}=-B^{c}\epsilon_{ij}x_j/2.$ Thus, we obtain the expression
of $\widetilde {p}_i$ by $p_i$ and $x_j$ in Eq.(\ref{Eq:tilde-p,x}),
there the momentum-momentum noncommutative parameter
\begin{equation}
\label{Eq:eta}
\eta_c=\hbar qB^{c}\sim 10^{-65}kg^{2}\cdot m^{2}\cdot s^{-2},
\end{equation}
where fundamental physical constants are taken from~\cite{Mohr}.

{\bf The Physical principles of determining $\widetilde {x}_i$}.
In the system described by $\widetilde {p}_i$, the position $x_i$
should be replaced by $\widetilde {x}_i$ accordingly. To find its
expression, Heisenberg quantization in Eq.(\ref{Eq:xp}) and
Bose-Einstein {(\bf B-E)} statistics have to be Maintained in the
system $(\widetilde {p}_i, \widetilde {x}_i)$.

{\bf Conditions of Maintaining {\bf B-E} statistics}.
When state vector space of identical bosons is constructed by
generalizing one-particle quantum mechanics, for two dimensional
harmonic oscillator, the annihilation-creation operators
$(\widetilde {a}_i, \widetilde {a}_ i^\dagger)$ are:
\begin{equation}
\label{Eq:a}
\widetilde {a}_i^\dagger=\sqrt{\frac{\mu\omega}{2\hbar}}\left
(\widetilde {x}_i - \frac{i}{\mu\omega}\widetilde {p}_i\right), \;
(i=1,2).
\end{equation}
where $\mu$ is mass of the considered particle and
$\omega$ is its characteristic frequency~\cite{JZZ06}.

The closed and complete {\bf B-E} algebra of $\widetilde {a}_i$ and
$\widetilde {a}_i^\dagger$ are~\cite{JZZ04}
\begin{equation}
\label{Eq:a-a}
[\widetilde {a}_i,\widetilde {a}_j^\dagger]=\delta_{ij}+
i\xi^{-2}\mu\omega\theta_c\epsilon_{ij},
[\widetilde {a}_i,\widetilde {a}_j]=[\widetilde
{a}_i^\dagger,\widetilde {a}_j^\dagger]=0,\\
(i,j=1,2),
 %
\end{equation}
Eq.(\ref{Eq:tilde-p,x}) is a linear transformation between two sets
of phase space variables $(\widetilde {p}_i,\widetilde {x}_i)$ and
$(x_i,p_i)$. It is proved that except Eq.(\ref{Eq:tilde-p,x}) any
other type of such linear transformations cannot maintain both the
physical space Heisenberg-Weyl algebra Eq.(\ref{Eq:tilde-xp}) and
the physical space bosonic algebra Eq.(\ref{Eq:a-a})~\cite{JZZ06}.

To maintain {\bf B-E} statistics the condition is that $\widetilde
{a}_i^\dagger$ and $\widetilde {a}_j^\dagger$ should be commuting,
$[\widetilde {a}_i^\dagger, \widetilde {a}_j^\dagger]=0.$ This
leads to two results:\\
(i) The consistency expression of $\widetilde {x}_i$ by $x_i$ and
$p_j$ is obtained.
By such ( $\widetilde {p}_i, \widetilde {x}_i$) in
Eq.(\ref{Eq:tilde-p,x}), Heisenberg quantization is maintained in physical
space.\\
(ii) {\bf The position-position noncommutative parameter $\theta_c$
must be a non-zero constant}. Using (\ref{Eq:a}) we obtain
$[\widetilde {a}_1^\dagger, \widetilde {a}_2^\dagger]=i\xi\mu \omega
[\theta_c-\eta_c/(\mu \omega)^2]/(2\hbar).$
From $[\widetilde {a}_i^\dagger, \widetilde {a}_j^\dagger]=0$, it
follows that~\cite{note-1}
\begin{equation}
\label{Eq:theta-eta}
\theta_c=\eta_c/(\mu \omega)^2.
\end{equation}
If $\theta_c$ was zero,
$[\widetilde {a}_1^\dagger,\widetilde {a}_2^\dagger]=-i\mu \omega
[\eta_c/(\mu \omega)^2]/(2\hbar)\ne 0$,
thus {\bf B-E} statistics could not be maintained. Therefore, the
maintenance of {\bf B-E} statistics leads to that $\theta_c$ being a
non-zero constant,
\begin{equation}
\label{Eq:non-zero theta}
\theta_c\ne 0.
\end{equation}
Although Eq.(\ref{Eq:non-zero theta}) is obtained in an example, it
is enough to ascertain that the existence of the non-zero $\theta_c$
is of universal significance.

 %
 %
Eq.(\ref{Eq:non-zero theta}) clarifies that the case 2) is excluded
by maintaining {\bf B-E} statistics. It concludes that {\bf physical
space is noncommutative space} (the case $[\widetilde {x}_i,\widetilde
{x}_j]=0$ cannot exist.)
Eq.(\ref{Eq:tilde-xp}) is the basis of physical
space quantum mechanics ({\bf PSQM}).

\vspace{0.3cm}

{\bf The position-position minimal uncertainty} of {\bf GHQ}. In
{\bf GHQ} of Eq.(\ref{Eq:tilde-xp}) there are two new minimal
uncertainties: The second equation in Eq.(\ref{Eq:tilde-xp})
indicates that the momentum-momentum minimal uncertainty
$(\Delta{\tilde p})_{min}\equiv (\Delta{\tilde
p}_1)_{min}=(\Delta{\tilde p}_2)_{min}$ of ${\bf {\tilde p}}$
reads (Neglect the high order term $\eta_c\theta_c$, thus $\xi\sim
1$)
\begin{equation}
\label{Eq:p-mini}
(\Delta{\tilde p})_{min}= \sqrt{\eta_c/2};
\end{equation}
From the third equation in Eq.(\ref{Eq:tilde-xp}), it follows the
position-position minimal uncertainty\\
$(\Delta{\tilde x})_{min}\equiv (\Delta{\tilde x}_1)_{min}
=(\Delta{\tilde x}_2)_{min}$ of $\bf {\tilde {x}}$ is
\begin{equation}
\label{Eq:x-mini}
(\Delta{\tilde x})_{min}= \sqrt{\theta_c/2}.
\end{equation}

$(\Delta{\tilde x})_{min}$ provides a fundamental minimal length
scale $(\Delta{\tilde x})_{min}$,
which leads to new space structures in noncommutative space.

The existence of the non-zero minimal length scale $(\Delta{\tilde
x})_{min}$ in Eq.(\ref{Eq:x-mini}) indicates that
the procedure of dividing a finite area in the $\tilde x_1-\tilde
x_2$ plane is not infinitive, it stopes at a minimal area, in which
lengths of any two orthogonal dimensions are not less than the
minimal length scale. 
 %
 %
Generalizing to three dimensional system, the procedure of dividing
a finite volume of space is not infinitive, it stopes at a minimal
volume, in which lengths of any three orthogonal dimensions are not
less than the minimal length scale.

\vspace{0.3cm}

{\bf Summary and Discussions.} New finds in this paper are:
(i) Heisenberg quantization Eq.(\ref{Eq:xp}) and {\bf HQM} were
proposed in empty space, but empty space does not exist.

(ii) Based on established experiments, {\bf GHQ} is obtained, and
{\bf PSQM} based on {\bf GHQ} has not free
parameters~\cite{note-2},\citer{Carr,Bert}.

(iii) By using Eq.(\ref{Eq:tilde-p,x}), all calculations of {\bf
PSQM} are realized by using variables ($x, p$). In the zero order of
$\eta_c$ and $\theta_c$, {\bf PSQM} recovers all predictions of {\bf
HQM}. First orders of $O(\eta_c)$ and $O(\theta_c)$ give
perturbation corrections of {\bf PSQM} to {\bf HQM}. Furthermore,
{\bf GHQ} and {\bf PSQM} include important new physics.

(iv) It explores the existence of a non-zero position-position
noncommutative parameter $\theta_c$ in Eq.(\ref{Eq:non-zero theta}),
and the relevant non-zero minimal length scale in
Eq.(\ref{Eq:x-mini}). The existence of such a minimal length scale
leads to essentially new space structure.

(v) It clarifies that physical space is noncommutative space. {\bf
GHQ} and {\bf PSQM} are quantum theories in noncommutative space.

(vi) {\bf The correlativity of the property of space with the motion
status}. In the special theory of relativity, Einstein found that
the length scale of space depends on the velocity of the inertial
frame in which measurements of the length scale are made. This is
the first discovery of the space property related to the motion
status.
 %
 %
Here, Eq.(\ref{Eq:theta-eta}) explores a new correlativity of the
property of space with the motion status of the system: space
non-commutativity is determined by the momentum non-commutativity.

\vspace{0.4cm}

The physical space algebra Eq.(\ref{Eq:tilde-xp}) is related to the
empty space algebra Eq.(\ref{Eq:xp}) by a similarity transformation.
However, Eq.(\ref{Eq:tilde-xp}) and Eq.(\ref{Eq:xp}) are
un-equivalency \cite{un-equiv}.
Because Eq.(\ref{Eq:xp}) and Eq.(\ref{Eq:tilde-xp}) are,
respectively, the foundations of {\bf HQM} and {\bf PSQM}, based on
such a un-equivalency, one expects essentially new physics emerged
from {\bf GHQ} and {\bf PSQM}. So the above new physics can be
understood.

The existence of a minimal length scale leads to that the space
structure of physical space is essentially different from one of
empty space. From the existence of the minimal length scale, it
follows the existence of the minimal space volume. It indicates that
point particle does not exist, any massive particle, including the
electron \cite{point model},~\citer{Rega,Cair}, has a minimal
volume.

\vspace{0.6cm}

{\bf Acknowledgements}

This work has been supported by NNSFC (the National Natural Science
Foundation of China) for financial support under grant numbers:
10575037, 10074014, 19674014, 19274017, 1880126 and SEDF (the
Shanghai Education Development Foundation).
The author would like to thank Huashan Hospital very much for the
outstanding medical service. The harmonious atmosphere in Ward 27
and Ward 28 of Inpatient lets author possible to clarify physics of
the paper and to organize the draft during his stay in Huashan
Hospital.


\begin{thebibliography}{99}
\bibitem{note-1}
 %
$\theta_c$ is dynamical parameter dependent. In the Lagrangian
formalism the definition of the canonical momentum is dynamical
dependent, so the dynamical parameter dependence of $\theta_c$ can
be understood.
\bibitem{un-equiv}
 %
It note that the transform matrix is not a orthogonal real matrix.
Furthermore, it is proved that a unitary similarity transformation
which transforms Eq.(\ref{Eq:tilde-xp}) and Eq.(\ref{Eq:xp}) to each
other does not exist. These elucidate the un-equivalency between the
two algebras~\cite{JZZ06}.
\bibitem{point model}
 %
For the electron, a non-zero electron Electric Dipole Moment has not
yet been found in all experiments to date, it only set improved
upper limit~\citer{Rega,Cair}. It seems that, up to now, experiments
are in favor of the view that the electron is size-less. However,
from the existence of the non-zero minimal length scale in physical
space, it follows that in the reality, the electron has a finite
volume, it is not a size-less point particle, even though one does
not know yet how small the electron is.
\end{thebibliography}
\end{document}